\documentclass[aps,prd,reprint,preprintnumbers,floats,epsfig,nofootinbib,amssymb,
nofootinbib]{revtex4}

\usepackage{slashed}
\usepackage{graphicx,color}
\usepackage{epsfig}
\usepackage{subfigure}
\usepackage{epsfig}
\usepackage{epstopdf}
\usepackage{dcolumn}
\usepackage{bm}
\usepackage{color}
\usepackage{ulem}
\usepackage{enumitem}
\usepackage[colorlinks,
linkcolor=black,
anchorcolor=black,
citecolor=black
]{hyperref}

\begin{document}

\baselineskip=15pt

\title{Dark photon kinetic mixing effects for the CDF W-mass measurement}

\author{Yu Cheng$^{1}$\footnote{chengyu@sjtu.edu.cn}}
\author{Xiao-Gang He${}^{1,2}$\footnote{hexg@sjtu.edu.cn}}
\author{Fei Huang${}^{1}$\footnote{fhuang@sjtu.edu.cn}}
\author{Jin Sun$^{1}$\footnote{019072910096@sjtu.edu.cn}}
\author{Zhi-Peng Xing$^{1}$\footnote{zpxing@sjtu.edu.cn}}

\affiliation{${}^{1}$Tsung-Dao Lee Institute, and School of Physics and Astronomy, Shanghai Jiao Tong University, Shanghai 200240, China}
\affiliation{${}^{2}$National Center for Theoretical Sciences, and Department of Physics, National Taiwan University, Taipei 10617, Taiwan}

\date{\today}

\vskip 1cm
\begin{abstract}
A new $U(1)_X$ gauge boson $X$ primarily interacting with a dark sector  can have renormalizable kinetic mixing with the standard model (SM) $U(1)_Y$ gauge boson $Y$. This mixing besides introduces interactions of dark photon and dark sector with SM particles, it also  modifies interactions among SM particles. The modified interactions can be casted into the oblique $S$, $T$ and $U$ parameters. We find that with the dark photon mass larger than the $Z$ boson mass, the kinetic mixing effects can reduce the tension of the W mass excess problem reported recently by CDF from $7\sigma$ deviation to within $3 \sigma$ compared with theory prediction. If there is non-abelian kinetic mixing between $U(1)_X$ and $SU(2)_L$ gauge bosons, in simple renormalizable models of this type a triplet Higgs is required to generate the mixing. We find that this triplet with a vacuum expectation value of order 5 GeV can naturally explain the W mass excess. 

\end{abstract}

\pacs{PACS numbers: }

\maketitle
    

\newpage

\section{Introduction}
Recently CDF collaboration announced their new measurement of W boson mass with a value of~\cite{CDF} $80,433.5 \pm 9.4 $ MeV which is 7$\sigma$ above the standard model (SM) prediction~\cite{pdg} of $80,357 \pm 6 $ MeV. This is a significant indication of new physics beyond the SM. A lot of efforts have been made to provide an explanation for this excess. Needless to say that better understanding of SM calculations, and also further experimental measurements are needed, nevertheless a lot of new ideas beyond the SM have merged to explain the W mass excess~\cite{Lu:2022bgw,Athron:2022qpo,Strumia:2022qkt,Yang:2022gvz,deBlas:2022hdk,Tang:2022pxh,Cacciapaglia:2022xih,Blennow:2022yfm,Zhu:2022scj,Sakurai:2022hwh,Fan:2022yly,Lee:2022nqz,yucheng,Song:2022xts,Bagnaschi:2022whn,Paul:2022dds,Bahl:2022xzi,Asadi:2022xiy,DiLuzio:2022xns,Athron:2022isz,Gu:2022htv,Heckman:2022the,Babu:2022pdn,Yuan:2022cpw,Du:2022pbp,Zhang:2022nnh,Du:2022fqv,Borah:2022zim,Baek:2022agi,Ghorbani:2022vtv,Zeng:2022lkk,Barrie:2022cub,Borah:2022obi,Chowdhury:2022moc,Arcadi:2022dmt,Popov:2022ldh,Carpenter:2022oyg,Nagao:2022oin,Kawamura:2022uft,Ghoshal:2022vzo,Perez:2022uil,Han:2022juu,Endo:2022kiw,Balkin:2022glu,Biekotter:2022abc,Krasnikov:2022xsi,Liu:2022jdq}. 
Although the CDF measurement is not fully consistent with LHC measurements ($m_W (\mbox{ATLAS})=80,370\pm 19$ MeV~\cite{ATLAS:2017rzl}, $m_W (\mbox{LHCb})=80,354\pm 32$ MeV~\cite{LHCb:2021bjt}), it may be viewed as a tantalizing evidence for new physics beyond the SM. 
In this work we study effects of a class of  well motivated dark photon models on the CDF W mass measurement. 

A dark photon $X_\mu$ from a $U(1)_X$ gauge group primarily coupling to a dark sector can have kinetic mixing with the SM gauge boson.  The kinetic mixing besides introduces interactions of dark photon and dark sector with SM particles, it also  modifies interactions among SM particles which can be tested to high precision data obtained by various experiments. It has long been realized that a dark photon $X_\mu$ can mix with the $U(1)_Y$ gauge boson $Y_\mu$ in the SM gauge group $SU(3)_C \times  SU(2)_L \times U(1)_Y$ through a renormalizable kinetic mixing term~\cite{Okun:1982xi, manohar, holdom, foothe}, $X^{\mu\nu} Y_{\mu\nu}$. Here $A^{\mu\nu} = \partial^\mu A^\nu - \partial^\nu A^\mu$. The phenomenological implications of this simple kinetic mixing have been studied extensively~\cite{Curtin:2014cca, search-he0, search-he1,search-he2, Kors:2004dx}. The kinetic mixing of the dark photon with the non-abelian gauge boson $W^a_\mu$, which transforms under the $SU(2)_L$ as a triplet represented by the superscript index ``a'',  has also been studied~\cite{non-abelian0, non-abelian1,non-abelian2,non-abelian3, non-abelian4,non-abelian5}. It turns out that  this requires additional efforts because the simple naive kinetic mixing $X^{\mu\nu} W^a_{\mu\nu}$ term is not gauged invariant. One needs to introduce a scalar type of entity transforming also as a triplet to make the relevant term gauge invariant. The simplest  one of such a entity is a scalar triplet $\Sigma^a$ transforming as $(1,\;3,\;0)$ under the $SU(3)_C\times SU(2)_L \times U(1)_Y$, with a non-zero vacuum expectation value (vev) $<\Sigma^0> = v_\Sigma$. Renormalizable models have been constructed recently~\cite{non-abelian4}. This type of model has some new interesting features, in particular CP violating kinetic mixing can also exist with testable consequences. 

Both types of models mentioned above will modify the interactions of the SM particles and therefore produce deviations from the SM predictions which can be tested by experimental data.
We find that the modified interactions can be casted into the oblique parameters $S$, $T$ and $U$ within the allowed parameter space, the kinetic mixing effects can help to explain the W mass excess of the recent CDF measurement. 
In the case of non-abelian kinetic mixing between $U(1)_X$ and $SU(2)_L$ gauge bosons, there are additional contributions to  the W mass excess besides the kinetic mixing effects due to the vev of triplet Higgs  required to generate the kinetic mixing. The triplet with a vev of order 5 GeV can naturally explain the W mass excess. We provide some details in the following.

\section{$S$, $T$, $U$ parameters in abelian kinetic mixing models}

With the kinetic mixing for the case of $U(1)_X\times U(1)_Y$, the kinetic terms of the bare fields $\tilde X$ and $\tilde Y$ and their interactions with other particles can be written as 
\begin{eqnarray}
\mathcal{L} = -{1\over 4} \tilde X_{\mu\nu} \tilde X^{\mu\nu} - {\sigma \over 2} \tilde X_{\mu\nu} \tilde Y^{\mu\nu} - {1\over 4} \tilde Y_{\mu\nu} \tilde Y^{\mu\nu} + j^\mu_Y \tilde Y_\mu + j^\mu_X \tilde X_\mu\;.
\end{eqnarray}
Here $j^\mu_X$ and $j^\mu_Y$ denote interaction currents of gauge fields $\tilde X$ and $\tilde Y$, respectively. The parameter $\sigma$ indicates the strength of the kinetic mixing.

After electroweak symmetry breaking, $\tilde Y$ and the neutral component of the $SU(2)_L$ gauge field $\tilde W^3$ can be written in the combinations of the ordinary SM photon field $\tilde A$ and the $Z$ boson field $\tilde Z$ as follows
\begin{eqnarray}
\tilde Y_\mu = \tilde c_W \tilde A_\mu -\tilde s_W \tilde Z_\mu\;,\;\; \tilde W^3_\mu = \tilde s_W \tilde A_\mu + \tilde c_W \tilde Z_\mu\;,
\end{eqnarray}
where $\tilde c_W \equiv \cos\tilde \theta_W$ and $\tilde s_W \equiv \sin \tilde \theta_W$ with $\theta_W$ being the weak mixing angle. Meanwhile, the $\tilde Z$ field receives a mass $ m_Z$.

The general Lagrangian that describes $\tilde A$, $\tilde Z$ and $\tilde X$ fields kinetic energy, and their interactions with the electromagnetic  current $j^\mu_{em}$, neutral $Z$-boson current $j^\mu_Z$ and dark current $j^\mu_X$ is given by~\cite{search-he2}.
\begin{eqnarray}
\mathcal{L} = &&-{1\over 4} \tilde X_{\mu\nu} \tilde X^{\mu\nu} - {1\over 4} \tilde A_{\mu\nu} \tilde A^{\mu\nu} - {1\over 4} \tilde Z_{\mu\nu} \tilde Z^{\mu\nu} 
-{1\over 2} {\sigma \tilde c_W}\tilde X_{\mu\nu} \tilde A^{\mu\nu} + {1\over 2} {\sigma \tilde s_W} \tilde X_{\mu\nu} \tilde Z^{\mu\nu}\nonumber\\
&& + j^\mu_{em} \tilde A_\mu + j^\mu_Z \tilde Z_\mu + j^\mu_X \tilde X_\mu +{1\over 2}m^2_Z \tilde Z_\mu \tilde Z^\mu\;,
\end{eqnarray}
where the $Z$ boson mass term is included. Here the currents for fermions with charge $Q_f$ and weak isospin $I^f_3$ in the SM are given by
\begin{eqnarray}
&&j^\mu_{em} = - \sum_f \tilde e Q_f \bar f\gamma^\mu f\;,\;\;j^\mu_Z = -{\tilde e \over 2 \tilde s_W \tilde c_W} \bar f\gamma^\mu ( g^f_V - g^f_A \gamma_5) f\;,\nonumber\\
&& g^f_V = I^f_3 - 2 Q_f \tilde s^2_W\;,\;\;\;\; g^f_A = I^f_3\;.
\end{eqnarray}
Note that the W boson field and its interactions are not affected directly.

The dark photon may be also massive. There are two popular ways of generating dark photon mass which give rise to different phenomenology. One of them is the ``Higgs mechanism'', in which the $U(1)_X$ is broken by the vev of an SM singlet $S$, which is charged under $U(1)_X$. In this case, the mixing of Higgs doublet and the Higgs singlet offers the possibility of searching for dark photon at colliders in Higgs decays~\cite{Curtin:2014cca, search-he1, search-he2,Davoudiasl:2012ig,Davoudiasl:2012ag,Davoudiasl:2013aya}. 
In this case, the singlet scenario can not explain the CDF W mass as shown in Refs.~\cite{Sakurai:2022hwh,Papaefstathiou:2022oyi}.
The other one is the ``Stueckelberg mechanism''~\cite{Kors:2004dx} in which an axionic scalar was introduced to allow a mass for $\tilde X$ without breaking $U(1)_X$. In our later discussion our concern is that the dark photon is massive  regardless where it comes from. We need to include a  mass term $(1/2)m^2_X \tilde X_\mu \tilde X^\mu$ in our discussions.  

One can rewrite the Lagrangian to remove the kinetic mixing terms so that the gauge fields kinetic energy terms are in the canonical form. This way to do this is not unique as discussed in Ref.~\cite{search-he2}. We choose to work with redefining the gauge fields such that photon  has no interaction with $j^\mu_X$. In this case, one redefines the fields as the following
\begin{eqnarray}
&&\left (\begin{array}{l}
\tilde A\\
\tilde Z\\
\tilde X
\end{array}
\right )
=\left ( \begin{array}{ccc}
1&{-\sigma^2\tilde s_W \tilde c_W\over \sqrt{1-\sigma^2}\sqrt{1-\sigma^2 \tilde c^2_W}}&{-\sigma \tilde c_W\over \sqrt{1-\sigma^2  \tilde c^2_W}}\\
0&{\sqrt{1-\sigma^2 \tilde c^2_W}\over \sqrt{1-\sigma^2 }}&0\\
0&{\sigma \tilde s_W\over \sqrt{1-\sigma^2}\sqrt{1-\sigma^2 \tilde c^2_W}}&{1\over \sqrt{1-\sigma^2 \tilde c^2_W}}
\end{array}
\right )
\left (\begin{array}{l}
\tilde A^\prime\\
\tilde Z^\prime\\
\tilde X^\prime
\end{array}
\right )\;,
\end{eqnarray}
to obtain the Lagrangian,
\label{eq:case_b_azx}
\begin{eqnarray}
\mathcal{L} = &&-{1\over 4} \tilde X^\prime_{\mu\nu} \tilde X^{\prime \mu\nu} - {1\over 4} \tilde A^\prime_{\mu\nu} \tilde A^{\prime \mu\nu} - {1\over 4} \tilde Z^\prime_{\mu\nu} \tilde Z^{\prime \mu\nu} 
\nonumber\\
&& + j^\mu_{em} \left(\tilde A^\prime_\mu   - {\sigma^2 \tilde s_W \tilde c_W\over \sqrt{1-\sigma^2} \sqrt{1-\sigma^2 \tilde c^2_W}}\tilde Z^\prime_\mu -{\sigma \tilde c_W\over  \sqrt{1-\sigma^2 \tilde c^2_W}}\tilde X^\prime_\mu \right)\\
&& + j^\mu_Z \left( {\sqrt{1-\sigma^2 \tilde c^2_W}\over \sqrt{1-\sigma^2}} \tilde Z^\prime_\mu \right )+ j^\mu_X\left(  {\sigma \tilde s_W\over \sqrt{1-\sigma^2}\sqrt{1-\sigma^2 \tilde c^2_W}}\tilde Z^\prime_\mu +{1\over \sqrt{1-\sigma^2 \tilde c^2_W}} \tilde X^\prime_\mu \right)\nonumber\\
&&+{1\over 2} m^2_Z {1-\sigma^2 \tilde c^2_W\over 1-\sigma^2} \tilde Z^\prime_\mu \tilde Z^{\prime \mu}+{1\over 2} m^2_X \left(  {\sigma \tilde s_W\over \sqrt{1-\sigma^2}\sqrt{1-\sigma^2 \tilde c^2_W}}\tilde Z^\prime_\mu +{1\over \sqrt{1-\sigma^2 \tilde c^2_W}} \tilde X^\prime_\mu \right)^2\;.\nonumber
\end{eqnarray}
We see that  the field $\tilde A^\prime$ is
already the physical massless photon field $A$ without the need of further mass diagonalization. However, the $\tilde Z^\prime$ and $\tilde X^\prime$ are mixed states. One needs to diagonalize the mass matrix in $(\tilde Z^\prime, \tilde X^\prime)$ basis, 
\begin{eqnarray}
\left ( \begin{array}{cc}
{m^2_Z (1-\sigma^2 \tilde c^2_W)^2 + m^2_X \sigma^2 \tilde s^2_W\over (1-\sigma^2)(1-\sigma^2 \tilde c^2_W)}&{m^2_X \sigma \tilde s_W\over \sqrt{1-\sigma^2}(1-\sigma^2 \tilde c_W^2)}\\
{m^2_X \sigma \tilde s_W\over \sqrt{1-\sigma^2}(1-\sigma^2 \tilde c_W^2)}& {m^2_X\over 1-\sigma^2 \tilde c^2_W}
\end{array}
\right )\;.
\end{eqnarray}
To obtain the diagonalized fields $Z$ and $X$, we introduce the mixing angle as
\begin{eqnarray}
&&\left (\begin{array}{c}
Z\\
X
\end{array}
\right )
=
\left (\begin{array}{cc}
c_\theta&s_\theta\\
-s_\theta&c_\theta
\end{array}
\right )
\left (\begin{array}{c}
\tilde Z^\prime\\
\tilde X^\prime
\end{array}
\right )\;,
\end{eqnarray}
with $c_\theta = \cos\theta$, $s_\theta = \sin\theta$, and 
\begin{eqnarray}\label{theta}
\tan(2\theta) = {2m^2_X \sigma \tilde s_W\sqrt{1-\sigma^2} \over m^2_Z (1-\sigma^2 \tilde c^2_W)^2 - m^2_X [1- \sigma^2(1+\tilde s^2_W)]}\;.
\end{eqnarray}
The diagonal masses  $\bar m_Z^2 = m^2_Z (1+\tilde z)$ which defines the parameter $\tilde z$,
and $\bar m^2_X$ corresponding to $Z$ and $X$ are given, respectively,  by
\begin{eqnarray}
&&\bar m^2_Z = {m^2_Z (1-\sigma^2 \tilde c_W^2)^2 + m^2_X \sigma^2 \tilde s_W^2\over (1-\sigma^2)(1-\sigma^2 \tilde c^2_W)} c^2_\theta 
+ {m^2_X \over 1-\sigma^2 \tilde c^2_W} s^2_\theta + 2 s_\theta c_\theta {m^2_X \sigma \tilde s_W\over \sqrt{1-\sigma^2} (1-\sigma^2 \tilde c^2_W)}\;,\nonumber\\
&&\bar m^2_X = {m^2_Z (1-\sigma^2 \tilde c_W^2)^2 + m^2_X \sigma^2 \tilde s_W^2\over (1-\sigma^2)(1-\sigma^2 \tilde c^2_W)} s^2_\theta 
+ {m^2_X \over 1-\sigma^2 \tilde c^2_W} c^2_\theta - 2 s_\theta c_\theta {m^2_X \sigma \tilde s_W\over \sqrt{1-\sigma^2} (1-\sigma^2 \tilde c^2_W)}\;.
\end{eqnarray}
The resulting Lagrangian is given by
\begin{eqnarray}
\mathcal{L} = &&- {1\over 4} A_{\mu\nu} A^{ \mu\nu} - {1\over 4} Z_{\mu\nu} Z^{ \mu\nu} 
+{1\over 2} \bar m^2_Z Z^\mu Z_\mu \nonumber\\
&& + j^\mu_{em} A_\mu   - j^\mu_{em} \left({\sigma^2 \tilde s_W \tilde c_W\over \sqrt{1-\sigma^2} \sqrt{1-\sigma^2 \tilde c^2_W}}c_\theta  + {\sigma \tilde c_W\over  \sqrt{1-\sigma^2 \tilde c^2_W}}s_\theta  \right)Z_\mu 
 + j^\mu_Z {\sqrt{1-\sigma^2 \tilde c^2_W}\over \sqrt{1-\sigma^2}} c_\theta Z_\mu\nonumber\\
 && + j^\mu_X\left(  {\sigma \tilde s_W\over \sqrt{1-\sigma^2}\sqrt{1-\sigma^2 \tilde c^2_W}}c_\theta +{1\over \sqrt{1-\sigma^2 \tilde c^2_W}} s_\theta  \right)Z_\mu \nonumber\\
&&-{1\over 4} X_{\mu\nu} X^{ \mu\nu} + {1\over 2} \bar m^2_X X^\mu X_\mu + j^\mu_X\left(  - {\sigma \tilde s_W\over \sqrt{1-\sigma^2}\sqrt{1-\sigma^2 \tilde c^2_W}}s_\theta +{1\over \sqrt{1-\sigma^2 \tilde c^2_W}} c_\theta \right) X_\mu\;.\nonumber\\
&& + j^\mu_{em} \left({\sigma^2 \tilde s_W \tilde c_W\over \sqrt{1-\sigma^2} \sqrt{1-\sigma^2 \tilde c^2_W}}s_\theta  -{\sigma \tilde c_W\over  \sqrt{1-\sigma^2 \tilde c^2_W}}c_\theta \right) X_\mu - j^\mu_Z  {\sqrt{1-\sigma^2 \tilde c^2_W}\over \sqrt{1-\sigma^2}} s_\theta X_\mu\;.\label{dark-int}
\end{eqnarray}
The above kinetic mixing calculation is also shown in appendices of Refs.~\cite{Davoudiasl:2012ag,Davoudiasl:2013jma}. Our results are more general  than their ones due to the existence of terms  $\sigma^2$, which is important for the following oblique parameters.

If one just considers dark photon kinetic mixing effects on SM particles, the relevant terms are the first two lines in the above Lagrangian. The rest terms involving dark sectors will not be directly related. 
To compare with precision experimental data and address the W mass excess, we now recast the dark photon effects in terms of oblique parameters. We find that the oblique $S$, $T$ and $U$ parameters will be generated. 
The derivation for the oblique parameters can be arrived at by first writing the modifications to the SM Lagrangian in the following way~\cite{Burgess:1993qk}, 
\begin{eqnarray}
L = &&{1\over 2} (1 +z -C) m^2_Z Z^\mu Z_\mu + (1+w-B) m^2_W W^\mu W^\dagger_\mu\nonumber\\
&& + \left(1-{A\over 2}\right) j^\mu_{em} A_\mu + \left(1-{C\over 2}\right) (j^\mu_Z  + G j^\mu_{em}) Z_\mu + \left(\left(1-{B\over 2}\right) j^\mu_W W^+_\mu+h.c.\right)\;.
\end{eqnarray}
where $j^\mu_W = - (\tilde e/\sqrt{2} \tilde s_W) \bar f^u \gamma^\mu LV_{KM} f^d$. Normalizing the fields and charges to the physical ones, one obtains the relations
\begin{eqnarray}
&&\alpha S = 4 s^2_W  c^2_W (A-C) - 4 s_W c_W (c^2_W - s^2_W) G\;,\;\;\alpha T = w - z\;,\nonumber\\
&&\alpha U= 4 s^2_W(s^2_W A - B+ c^2_W C - 2 s_W c_W G)\;.
\end{eqnarray}
To the leading order, $\tilde s_W$ and $\tilde c_W$ can be replaced by $s_W$ and $c_W$ in the above  $C$ and $G$.
 
In our case, since there are no modifications to the $W^\pm_\mu$, whose coupling and bare mass, $B$ and $w$ are both zero.  
Also we see from Eq.~(\ref{dark-int}) that there exists no modification for photon interaction, therefore $A=0$. We obtain
\begin{eqnarray}
C=2\left(1-\frac{\sqrt{1-\sigma^2  c_W^2}}{\sqrt{1-\sigma^2}} c_\theta   \right)\;,\;\;\;\;G=-\frac{\sigma^2  s_W  c_W}{1-\sigma^2  c_W^2}-\frac{\sigma   c_W \sqrt{1-\sigma^2}}{1-\sigma^2  c_W^2}\frac{s_\theta}{c_\theta}\;,\;\;\;\; z=C+\tilde z\;.
\end{eqnarray}
Here the definition for $\tilde z$ is shown below Eq.~(\ref{theta}).
We obtain oblique parameters to the first order in $\sigma^2$ as
\begin{eqnarray}
	&&	\alpha S =\frac{ 4s_W^2 c_W^2 \sigma^2}{1-m_X^2/m_Z^2}\left(1-\frac{s_W^2}{1-m_X^2/m_Z^2} \right)\;,\nonumber \\
	&&		\alpha T =-\sigma^2 s_W^2 \frac{ m_X^2 /m_Z^2} {(1-m_X^2/m_Z^2)^2}\;,\nonumber \\
	&& 	\alpha U =4s_W^4 c_W^2 \sigma^2\left(-\frac{ 1-2m_X^2/m_Z^2}{(1-m_X^2/m_Z^2)^2}+ \frac{2}{1-m_X^2/m_Z^2} \right)\;.
	\label{stu}
\end{eqnarray}
The above leads to correction to the W mass as
\begin{eqnarray}
\Delta m_W^2&&=m_Z^2 c_W^2 \left(-\frac{\alpha S }{2(c_W^2-s_W^2)}+\frac{c_W^2 \alpha T}{(c_W^2-s_W^2)}+ \frac{\alpha U} {4s_W^2}\right)\nonumber\\
&&=- m_Z^2 c_W^2 
	 \frac{m_Z^2(1-s_W^2) \sigma^2 s_W^2}{(m_X^2-m_Z^2)(-1+2s_W^2)}\;. \label{Dwmass}
\end{eqnarray}
Other electroweak precision observables for the dark photon have been calculated as shown in Refs.~\cite{Hook:2010tw,Curtin:2014cca}.

\section{$S$, $T$,  $U$ parameters in non-abelian kinetic mixing models}

We now discuss how the W mass is modified in a class of non-abelian kinetic mixing models. 
This is the class of models in which kinetic mixing between the $U(1)_X$ gauge boson $\tilde X_\mu$ and $SU(2)_L$ gauge boson $\tilde W^a_\mu$ can be induced.  Here $\tilde W^a_\mu$ transforms as a $SU(2)_L$ triplet. To realize such kinetic mixing, the group index ``a'' needs to be balanced which can be achieved easily by introducing a scalar triplet $\Sigma^a$. With the help of $\Sigma^a$ the kinetic mixing terms of the following forms can be gauge invariant
\begin{eqnarray}
\tilde X^{\mu\nu} \tilde W^a_{\mu\nu} \Sigma^a\;, \;\;\;\;\epsilon^{\mu\nu \alpha\beta} \tilde X_{\mu\nu} \tilde W^a_{\alpha\beta} \Sigma^a\;.
\label{non-abelian}
\end{eqnarray}
The component fields of $\tilde W^a$ and $\Sigma^a$ are given by
\begin{eqnarray}
\sigma^a \tilde W^a_\mu = \left (\begin{array}{cc}  \tilde W^3_\mu&\; \sqrt{2} \tilde W^+_\mu\\ \sqrt{2}\tilde W^-_\mu&\; -\tilde W^3_\mu \end{array} \right )\;,\;\;\;\;
\sigma^a \Sigma^a = \left (\begin{array}{cc}   \Sigma^0&\; \sqrt{2}\Sigma^+\\ \sqrt{2} \Sigma^-&\; -\Sigma^0 \end{array} \right )\;. 
\end{eqnarray}
Here $\sigma^a$ are the Pauli matrices. When the $\Sigma^a$ neutral component ($\Sigma^0$) develops a non-zero vev $<\Sigma^0> = v_\Sigma$, the kinetic mixing in the usual form from the first term, $\sqrt{2}\tilde X^{\mu\nu} \tilde W^3_{\mu\nu} v_\Sigma$, and a new form from the second term, $\sqrt{2}\epsilon^{\mu\nu\alpha\beta}\tilde X_{\mu\nu} \tilde W^3_{\alpha\beta} v_\Sigma$ will be induced.

Note that had one replaced $\tilde W^3_{\alpha\beta}$  by $\tilde Y_{\alpha\beta}$, then  $\epsilon^{\mu\nu\alpha \beta} \tilde X_{\mu\nu} \tilde Y_{\alpha \beta} = 2 \partial^\mu (\tilde X^\nu \tilde Y_{\mu\nu})$ would be a total derivative, which would have no perturbative effects. The existence of  the term $W^-W^+$ in $\tilde W^3_{\mu\nu} = s_ W \tilde A_{\mu\nu} + c_W \tilde Z_{\mu\nu} +ig(W^-_\mu W^+_\nu - W^-_\nu W^+_\mu)$, will give rise an additional new term $2ig \epsilon^{\mu\nu\alpha\beta} \tilde X_{\mu\nu} W^-_\alpha W^+_\beta$, which cannot make $\epsilon^{\mu\nu\alpha\beta}\tilde X_{\mu\nu} \tilde W^3_{\alpha\beta}$ as a total derivative one again and has physical effects. Some interesting implications have been studied in Ref.~\cite{non-abelian4, non-abelian5}.

The operators in Eq.~(\ref{non-abelian}) are dimension 5 ones which is nonrenormalizable. If one insists on renormalizability of the model, additional ingredients need to be introduced to generate them at the loop level. A specific renormalizable model has been constructed recently~\cite{non-abelian5}. However, the kinetic mixing parameters  generated are too small~\cite{non-abelian5} to make a significant impact on W mass, and can be neglected.  But in this class of models, there are still two contributions which can affect the W mass significantly. One is the possible mixing term $-(1/2)\sigma \tilde X^{\mu\nu} \tilde Y_{\mu\nu}$ discussed earlier which generates the $S$, $T$, and $U$ parameters given in Eq.~(\ref{stu}). Another one is the non-zero $v_\Sigma$ of the triplet $\Sigma^a$ generated modification to the electroweak precision parameter $\rho$. We have~\cite{pdg, yucheng} 
\begin{eqnarray}
\rho = 1 + {4 v^2_\Sigma\over v^2} = 1 + \alpha T_\Sigma\;,
\end{eqnarray}
where $v = 246$ GeV is the SM Higgs vev.  
Here we only consider the tree-level contribution. If further considering the one-loop contribution, the additional term $O(\Delta m)$ with $\Delta m=m_{H^+}-m_H$  will emerge as shown in Ref.~\cite{Khan:2016sxm}.

The term  $T_\Sigma = 4 v^2_\Sigma/\alpha v^2$ is an addition to the $T$ parameter which needs to be considered in this class of model. Therefore for this class of models Eq.(\ref{Dwmass}) is modified to
\begin{eqnarray}
\Delta m_W^2&&=m_Z^2 c_W^2 \left(-\frac{\alpha S }{2(c_W^2-s_W^2)}+\frac{c_W^2 \alpha T}{(c_W^2-s_W^2)}+ \frac{\alpha U} {4s_W^2}\right)\nonumber\\
&&=- m_Z^2 c_W^2 
	 \frac{m_Z^2(1-s_W^2) \sigma^2 s_W^2}{(m_X^2-m_Z^2)(-1+2s_W^2)} + m^2_Z c^2_W {c^2_W\over c^2_W - s^2_W} {4 v^2_\Sigma\over v^2}\;. \label{mDwmass}
\end{eqnarray}

\section{numerical analysis and conclusions}

We are now ready to put things together to analyze whether the dark photon models can accommodate the W mass excess  indicated by the recent CDF result. The CDF result is 7$\sigma$ above the SM prediction, which implies that the new contributions must have $\Delta m^2_W >0$, so that $\Delta m^{CDF}_W =\sqrt{(m^{SM}_W)^2 + \Delta m^2_W} - m^{SM}_W \approx  \Delta m^2_W/(2m^{SM}_W)$ to produce the required $70$ MeV.

\begin{figure}[!t]
 \centering
 \subfigure[\label{sigma_mx}]
 {\includegraphics[width=.486\textwidth]{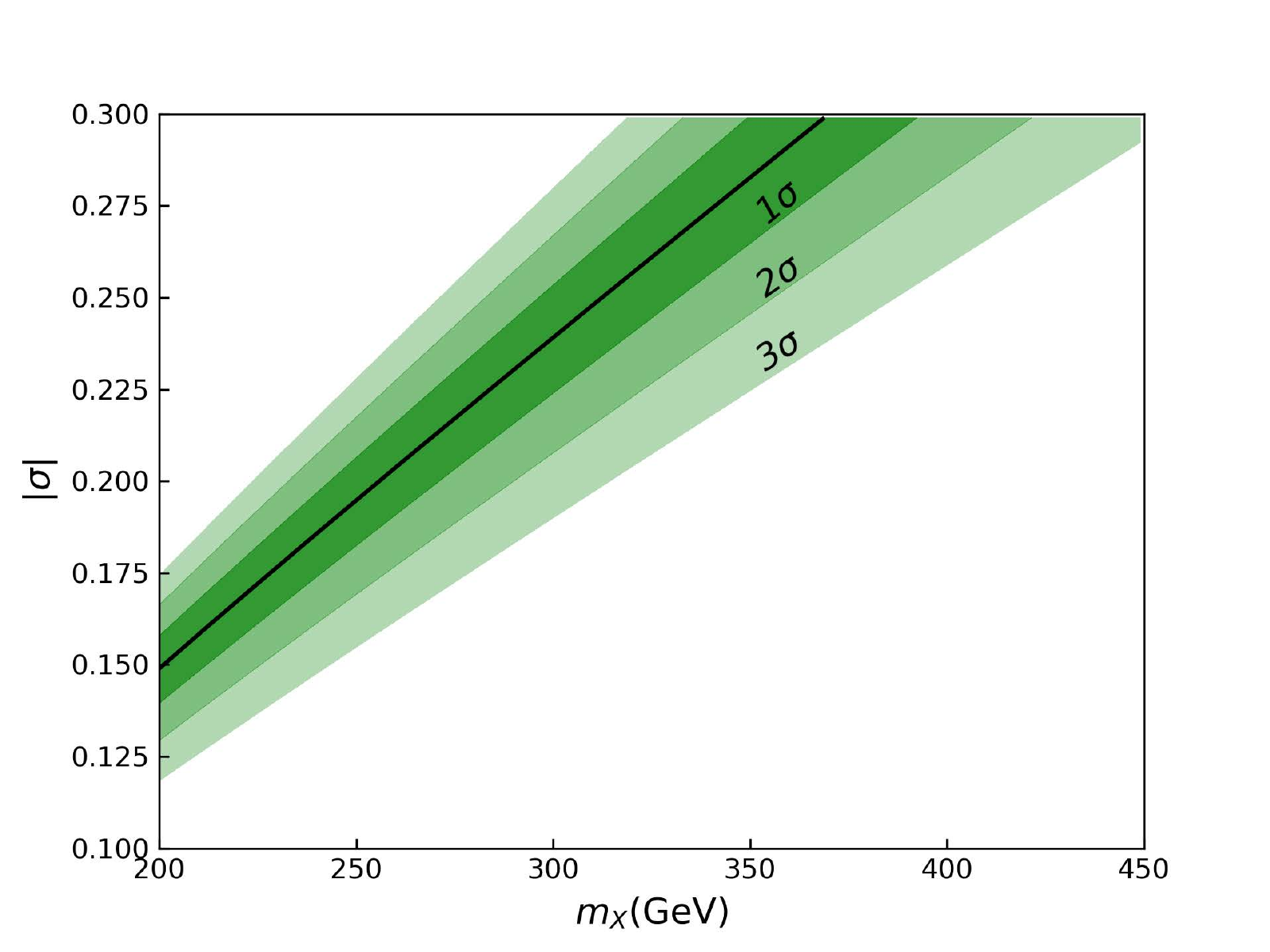}}
 \subfigure[\label{STU}]
 {\includegraphics[width=.486\textwidth]{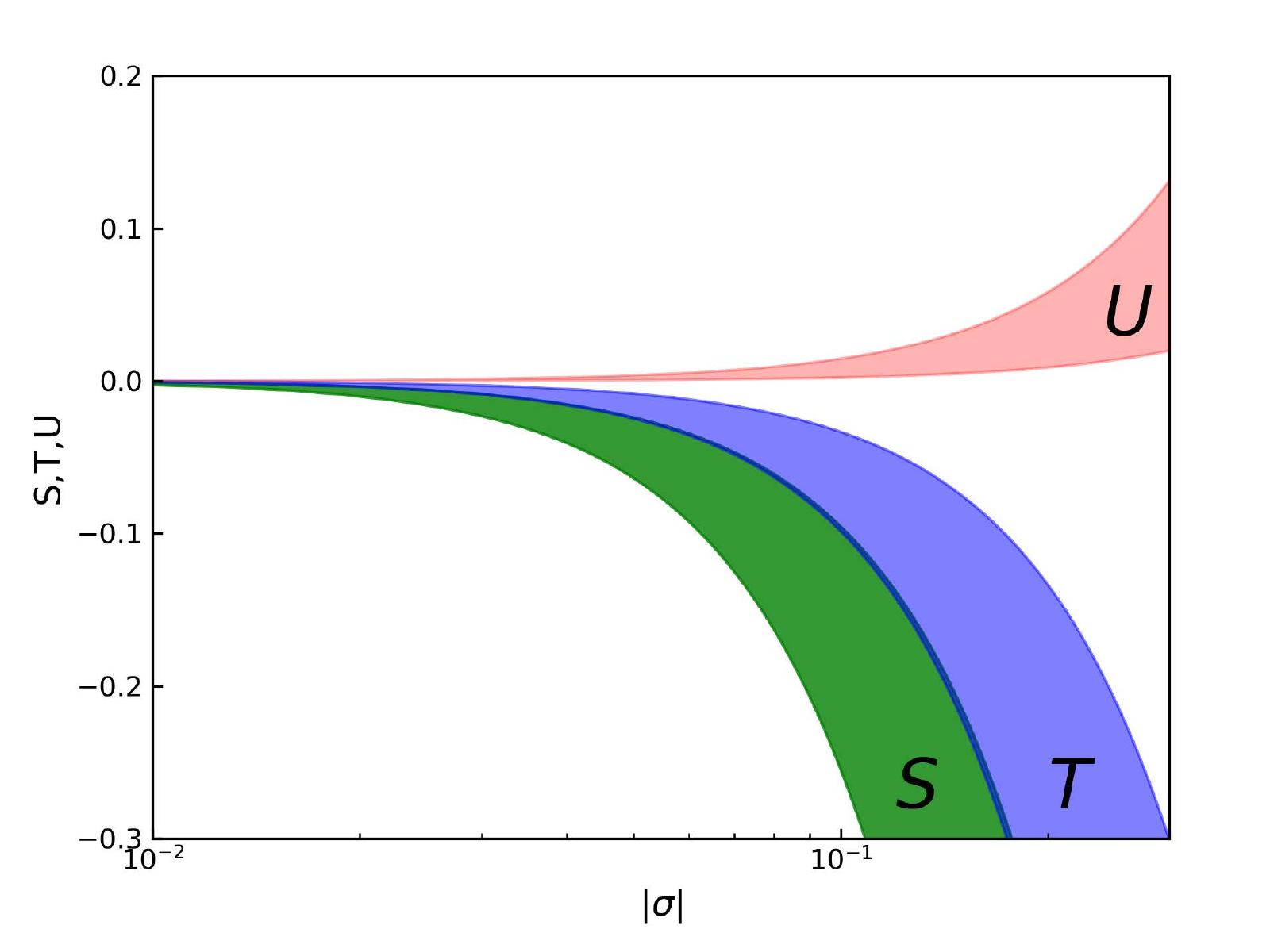}}
 \caption{ (a) The CDF allowed regions  in $m_{X} - |\sigma|$ plane. The allowed parameter space is shown in black line for central value, the $1\sigma$, $2\sigma$ and $3\sigma$ ranges are also shown. 
 	(b) The $S$, $T$ and $U$ parameters as functions of $|\sigma|$. The colored band means  the S,T,U ranges for fixing $m_X=(200, 300)$ GeV. For S and T,  the lower line and upper line mean $m_X=200$ GeV and $m_X=300$ GeV, respectively. For U, the situation is contrary. Thus, the size of parameters decrease when $m_X$ increases.  
 }
 \label{sigma}
\end{figure}

\begin{figure}[!t]
 \centering
 \subfigure[\label{vsigma_sigma}]
 {\includegraphics[width=.486\textwidth]{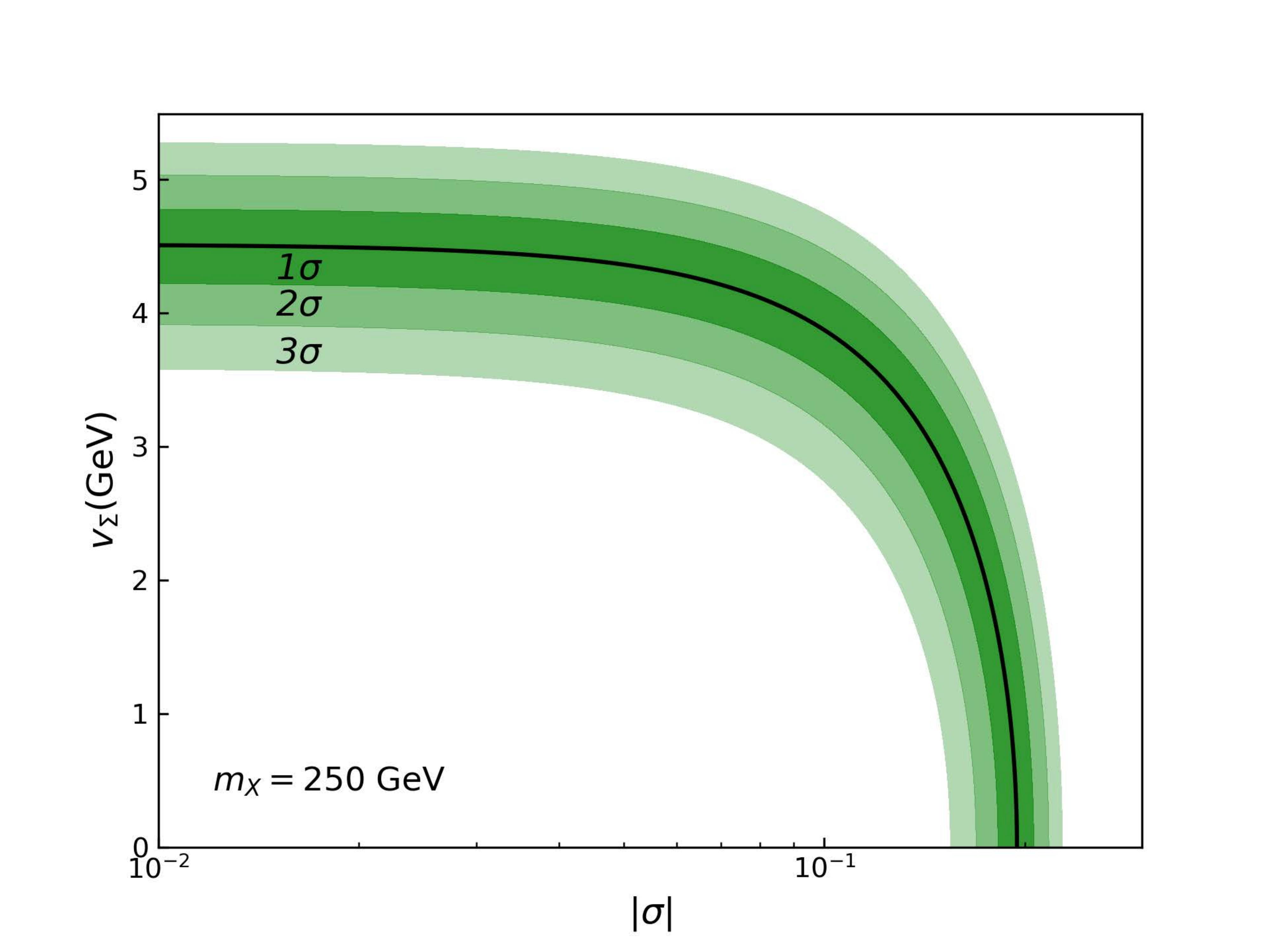}}
 \subfigure[\label{vsigmaT}]
 {\includegraphics[width=.486\textwidth]{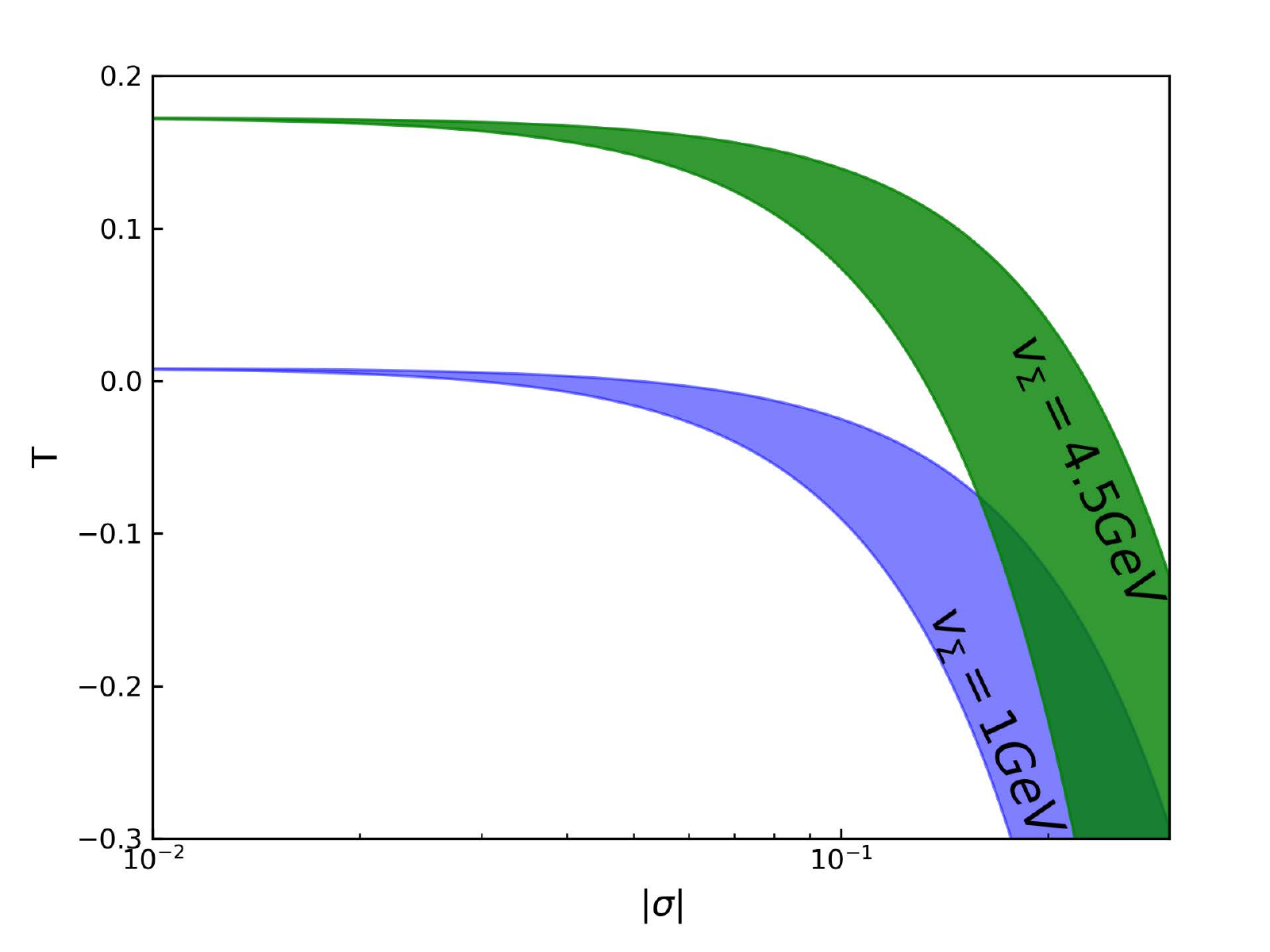}}
 \caption{  (a) The CDF allowed regions  in $|\sigma |- v_{\Sigma}$ plane for the fixed values  $m_X=250$ GeV. (b) The $T$ parameter for two different values of $v_\Sigma$.
}
 \label{STUconstraint}
\end{figure}

In the case of abelian kinetic mixing, we see from Eq.~(\ref{Dwmass}) that in order to have $\Delta m^2_W >0$, the dark photon mass $m_X$ must be larger than $m_Z$. We therefore confine our analysis in this range for $m_X$. When $m_X$ becomes larger, a larger $|\sigma|$ is needed. We plot the allowed ranges in Fig.\ref{sigma_mx}  for $m_X$ and $\sigma$ within  the central value, $1\sigma$, $2\sigma$ and $3\sigma$ boundaries, respectively.  We see that there are ranges in the $m_X-|\sigma|$ plane which can solve the W mass excess  problem. CMS has searched~\cite{CMS} for dark photon 
in the range of $m_X$ below about 200 GeV decaying to $\mu^+ \mu^-$ final states and gives a stringent constraint on $\sigma <10^{-3}-10^{-2}$. 
To evade the CMS constraint, we choose that $m_X>200$ GeV. Then we further find $|\sigma|>0.125$ to address the W mass excess problem from Fig.\ref{sigma_mx}.  Since our expansion parameter is $\sigma^2$ as shown in  Eq.~(\ref{Dwmass}), we consider $\sigma \sim 0.2- 0.3$  is a reasonable range. 
With improved high energy search similar to what carried out by CMS, the model can be more stringently constrained. 

Global fit of electroweak precision data, has given constraint on $S$, $T$, and $U$ separately.  Recently, the results of EW global fit with CDF W  mass obtain the oblique parameters: $S$, $T$, and $U$. 
The EW input parameters, such as $m_{W,Z}, \alpha_s, \Gamma_{Z,W}, A_f, A^{0,f}_{FB}, R_f, \sin^2\theta_{eff}$,  are in details shown in Ref.~\cite{Lu:2022bgw}.
In our evaluations, contributions to S, T, U parameters are calculated by using the Fermi constant $G_F$ and redefining $Z$ boson mass $\bar m_Z$ and the electromagnetic fine structure constant $\bar \alpha_{em}=\alpha_{em}/(1-\sigma^2 c_W^2)$~\cite{search-he2} as independent input parameters.
The fit values obtained from Ref.~\cite{Lu:2022bgw} are used for comparison: $S=0.06\pm0.1$, $T=0.11\pm0.12$, $U=0.14\pm0.09$. 
As shown in Fig.~\ref{STU}, it turns out that although we can obtain the CDF measured W mass, but it is not possible to satisfy the bounds on the $S$, $T$ and $U$ parameters within 2$\sigma$ allowed ranges. But within 3$\sigma$ allowed ranges, the abelian kinetic mixing effect can accommodate the CDF W mass measurement.

In the non-abelian kinetic mixing case, with the help of $v_\Sigma$ in the range of a few GeV, the model can easily accommodate the CDF W mass excess with  very small $\sigma$. We now discuss how a non-zero $v_\Sigma$ affects the model parameters. In this case, from Fig.\ref{vsigma_sigma} we see that a $v_\Sigma$ in the range of a few GeV can help solve the CDF W mass excess problem even with a very small kinetic mixing $\sigma$. The expressions for $S$ and $U$ are not changed compared with abelian kinetic mixing case, but the total $T$ needs to add an additional $T_\Sigma$  at the tree level. Therefore, the ranges for $S$ and $U$ will keep the same with Fig.\ref{STU}. Instead, $T$ will be modified depending on the $v_\Sigma$  as shown in Fig.\ref{vsigmaT}.
This result in changing the relative size of the parameters for a given $m_W$. Without $T_\Sigma$, $m_X$ cannot be too much larger than the CMS upper bound of $200$ GeV, and $\sigma$ cannot be much smallar than 0.2 or so. When one includes $T_\Sigma$ in the analysis, a much larger $m_X$ and also a smaller $\sigma$ can be allowed if the model is required to solve the W mass excess. 
In this case, the absolute values of $S$ and $U$ can be made small to satisfy the allowed global fit.  

Before summary, we would like to comment about a possible consequence of  the Z boson couples to dark sector. If the dark sector particles are enough light,  Z can decay into them to enhance the invisible width.  As an example, we assume that there is a vector-like fermion $f$ current $j^\mu_X = \tilde g \bar f \gamma^\mu f$ coupling to the original $\tilde X_\mu$. After normalizing the couplings and fields, we have
\begin{eqnarray}
L_{int}= \tilde g  \left(  {\sigma \tilde s_W\over \sqrt{1-\sigma^2}\sqrt{1-\sigma^2 \tilde c^2_W}}c_\theta +{1\over \sqrt{1-\sigma^2 \tilde c^2_W}} s_\theta  \right) \bar f \gamma^\mu f Z_\mu \approx   \tilde g {\sigma s_W m^2_Z\over m^2_Z - m^2_X}  \bar f \gamma^\mu f Z_\mu \;.
\end{eqnarray}
This interaction gives a invisible decay width for $Z \to f\bar f$
\begin{eqnarray}
\Gamma  = \frac{\tilde g^2}{12\pi}\frac{\sigma^2 s_W^2}{(1-m_X^2/m_Z^2)^2}m_Z \sqrt{1-\frac{4m_f^2} { m_Z^2}}\left(1+\frac{2m_f^2}{m_Z^2}\right)\;.
\end{eqnarray}
For the fermion with a very small  mass, if fixing $m_X=250$ GeV, and $\tilde g=g_Y=0.356$, we obtain the branching ratio as $2.8\times 10^{-5}(\sigma/0.2)^2(\tilde g/g_Y)^2$. Using the $Z$ decay width in Ref.~\cite{pdg}, one obtains $Br^{new}(Z \to invisible)=2.3 \times 10^{-3}$. The $Z$ invisible decay width agrees with SM prediction well. As long as $\sigma \tilde g$ is smaller than 0.65, one can safely satisfy the data.
 
To summarize, we have studied the recent CDF measurement of W mass on two classes of dark photon models, one is the abelian kinetic mixing case due to a dark photon abelian $U(1)_X$ and SM $U(1)_Y$  gauge boson mixing,  and another one is the non-abelian kinetic mixing from a dark photon $U(1)_X$ and another non-abelian SM $SU(2)_L$ gauge boson mixing. 
This mixing besides introduces interactions of dark photon and dark sector with SM particles, it also  modifies interactions among SM particles. We recast these modifications into the well know oblique $S$, $T$ and $U$ parameters. We find that with the dark photon mass larger than the $Z$ boson mass, the kinetic mixing effects can reduce the  tension of the W mass excess problem from $7\sigma$ to within $3 \sigma$ compared with theory prediction. If there is non-abelian kinetic mixing between $U(1)_X$ and $SU(2)_L$ gauge bosons, in simple renormalizable models of this type a triplet Higgs is required to generate the mixing. We find that this triplet with a vacuum expectation value of order 5 GeV can naturally explain the W mass excess.


\begin{acknowledgments}

This work was supported in part by Key Laboratory for Particle Physics, Astrophysics and Cosmology, Ministry of Education, and Shanghai Key Laboratory for Particle Physics and Cosmology (Grant No. 15DZ2272100), and in part by the NSFC (Grant Nos. 11735010, 11975149, and 12090064). XGH was supported in part by the MOST (Grant No. MOST 109-2112-M-002-017-MY3). ZPX was supported  by the NSFC (Nos. 12147147).\\
\end{acknowledgments}
{\bf Note added}
\\

 After our paper appeared,  Holdom  brought us the attention of Ref.~\cite{Holdom:1990xp} where the same S, T, U parameters had been calculated. Our expressions of S, T, U parameters  agree with each other.


\end{document}